\documentclass{kluwer}    
\usepackage[dvips]{graphicx}
\newdisplay{guess}{Conjecture}

\newcommand{\etal}{{\it et al.\ }}

\newcommand{\gta}{\stackrel{>}{\scriptstyle\sim}}

\begin{document}
\begin{article}
\begin{opening}
\title{Chemically Consistent Evolutionary Synthesis}
\author{Uta Fritze -- v. Alvensleben, Peter Weilbacher, Jens Bicker}
\runningauthor{Fritze--v. Alvensleben, Weilbacher \& Bicker}
\runningtitle{Chemically Consistent Evolutionary Synthesis}
\institute{Universit\"ats-Sternwarte G\"ottingen, Geismarlandstr. 11, 37083
G\"ottingen, Germany}
\begin{abstract}
To account for the range of stellar metallicities in local galaxies and for the increasing importance of low metallicities at higher redshift we present chemically consistent models for the spectral 
and chemical evolution of galaxies over cosmological timescales. 
We discuss advantages, limitations and future prospects of our approach. 
\end{abstract}
\end{opening}

\section{Introduction}
Any stellar system with a star formation history ({\bf SFH}) more extended 
than a massive star's lifetime (plus the cooling time of the gas) will be composite not only in age but also in metallicity. Indeed, broad metallicity distributions are observed in galaxies throughout (e.g. Rocha-Pinto \& Maciel 1998, Carollo \& Danziger 1994, Grebel 1997). With few exceptions stellar as well as gaseous metallicities are subsolar even in the local Universe and, of course, more so in the early Universe (e.g. Ferguson \etal 1998, van Zee \etal 1998, Loewenstein 1999, Lowenthal \etal 1997, Trager \etal 1997, Pettini et al. 2001, Mehlert et al. 2002, Pettini et al. 2002). 
 
\section{Chemically Consistent Modelling}
Evolutionary synthesis starts from a gas cloud of primordial abundances that initially comprises the total mass of the galaxy, give a star formation history ${\rm \Psi(t)}$ and an IMF. The 1$^{st}$ generation of stars is formed with ${\rm Z = 0}$. We solve a modified form of Tinsley's equations with stellar yields (SNII, SNI, PN, and stellar mass loss) and lifetimes for ${\rm Z = 0}$ to obtain the time evolution of the ISM abundances of individual elements $^{12}$C, ..., $^{56}$Fe. The next generation of stars is formed with abundances $Z>0$, and again, the modified Tinsley equations are to be solved with stellar yields and lifetimes for $Z>0$, ... . Via the SFH galaxy evolution and stellar evolution become intimately coupled. Different SFHs -- short and burst-like vs. mild and $\sim$ constant -- lead to different abundance ratios between elements with different nucleosynthetic origin, as e.g. [C/O], [N/O], or [$\alpha$/Fe]. C and N originate in intermediate mass stars, Fe has important SNIa contributions, both leading to a delayed production with respect to the SNII products O, Mg, etc. A threshold in metallicity, below which SNIa may be inhibited (Kobayashi \etal 1998), would further enhance this effect. In principle, stellar evolutionary tracks and yields are required not only for various metallicities (and He contents), but also for various abundance ratios. However, no complete grid of stellar evolutionary tracks or yields for varying abundance ratios [${\rm C,~N,~\alpha/Fe}$] is available yet. {\bf Our chemically consistent ($=$ {\bf cc}) evolutionary synthesis models follow the evolution of ISM abundances together with the spectrophotometric properties of galaxies and account for the increasing initial metallicity of successive generations of stars.}

\noindent
We use various sets of input physics -- stellar evolutionary isochrones, model atmosphere spectra, yields, lifetimes, and remnant masses -- ranging in metallicity from ${\rm Z=4 \cdot 10^{-4}}$ to ${\rm Z=0.05}$. Our cc models simultaneously describe the spectrophotometric {\bf and} the chemical evolution, both as a function of time and -- for any cosmological model as given by ${\rm H_0, \Omega_0, \Lambda_0,}$ and a redshift of galaxy formation ${\rm z_f}$ --  as a function of redshift. The chemo-cosmological evolution only needs a transformation (time t $~\rightarrow~$ redshift z). The redshift evolution of apparent magnitudes UBVRIJHK and colors requires evolutionary corrections ${\rm e_{\lambda}}$ (since a galaxy at ${\rm z > 0}$ is younger) and cosmological corrections ${\rm k_{\lambda}}$ (since its spectrum is redshifted).

\noindent
The two basic parameters of evolutionary synthesis models are the IMF, for which we use Salpeter, and SFHs ${\rm \Psi(t)}$, appropriate for different spectral galaxy types. SFHs are determined from the requirement that models after a Hubble time simultaneously match many observed average properties of the respective galaxy types like colors, template spectra (UV ... K), gas content and ISM abundances. We caution that our models are simple 1-zone descriptions without dynamics or spatial resolution, meant to describe global average
quantities like integrated spectra or colors, and HII region abundances around ${\rm \sim 1~R_e}$. They only have one gas phase, account for the lifetimes of the stars before they restore recycled gas but neglect its cooling time. Being closed-box models at the present stage, they should not be used for dwarf or strong starburst galaxies with large-scale in- and outflows. 

\noindent
The chemical aspects of our cc models were presented and compared to observations in Lindner \etal (1999), the cc spectral evolution using Geneva stellar evolutionary tracks by Fritze - v. A. (1999) and M\"oller \etal (2001). 

\section{Results and Comparison with Solar Metallicity Models}
Here we use the most recent and complete Padova isochrones including the thermal pulsing AGB phase for stars of ${\rm 0.9-8~M_{\odot}}$ (see Schulz \etal 2002 for details). 
Fig. 1 shows the decomposition of the total light emitted by 12 Gyr old E and Sb models in three filter bands U, V, K into the relative contributions of stars from 5 different metallicity subpopulations. Note the very broad stellar metallicity distribution in the E model, the lower contribution of metal-poor stars and the absence of ${\rm Z=0.05}$ stars in the Sb model. 

\begin{figure}[ht]
\includegraphics[width=0.5\columnwidth]{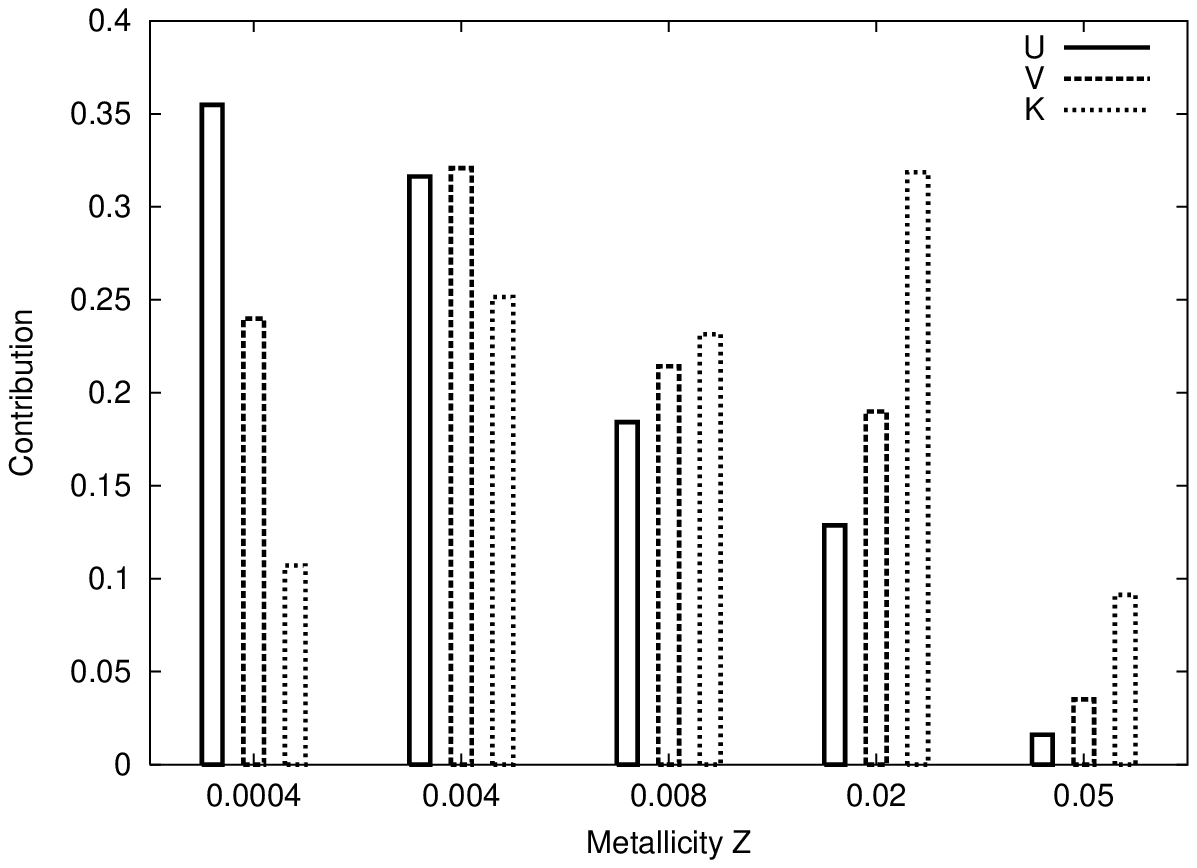}
\includegraphics[width=0.5\columnwidth]{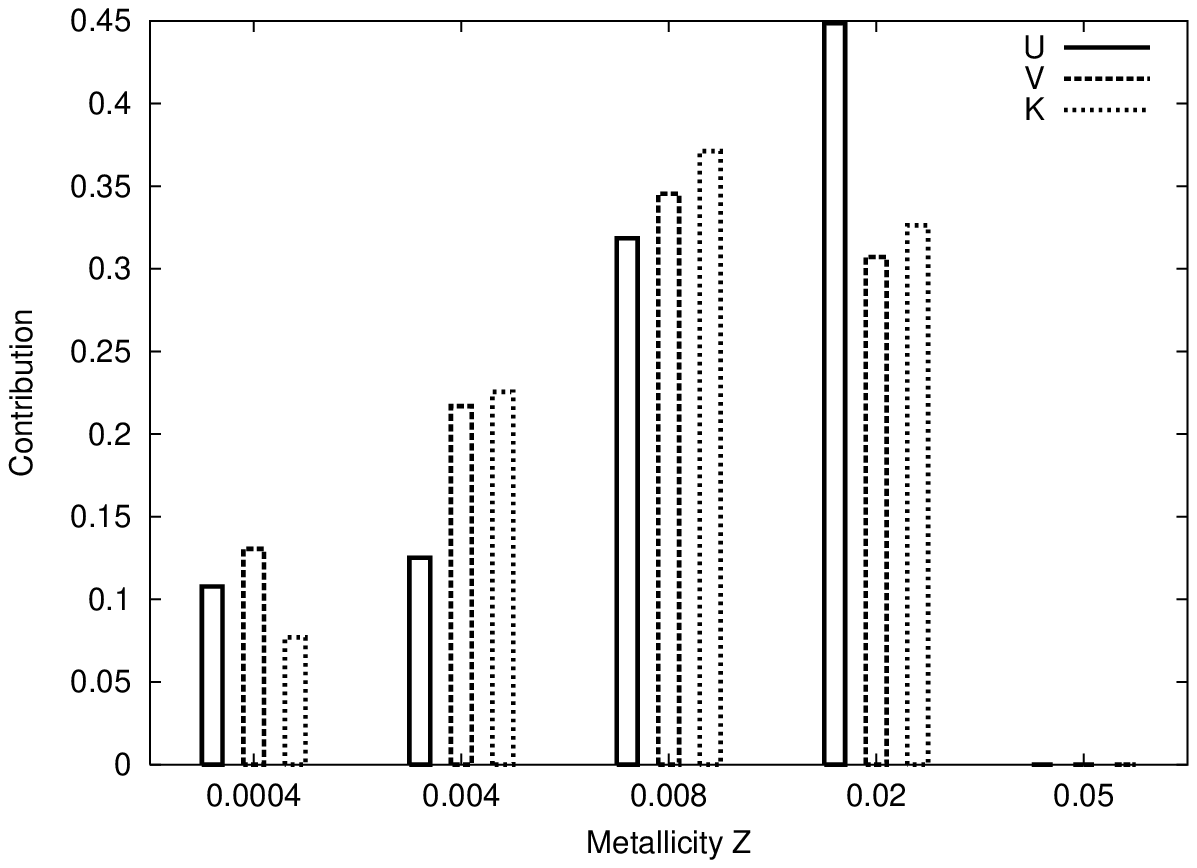}
\caption{Decomposition of the U-, V-, and K-light of 12 Gyr old E and Sb galaxies into the relative contributions from stellar subpopulations of different metallicities.}
\end{figure}

\vspace{-0.5cm}
\begin{figure}[ht]
\includegraphics[width=0.5\columnwidth]{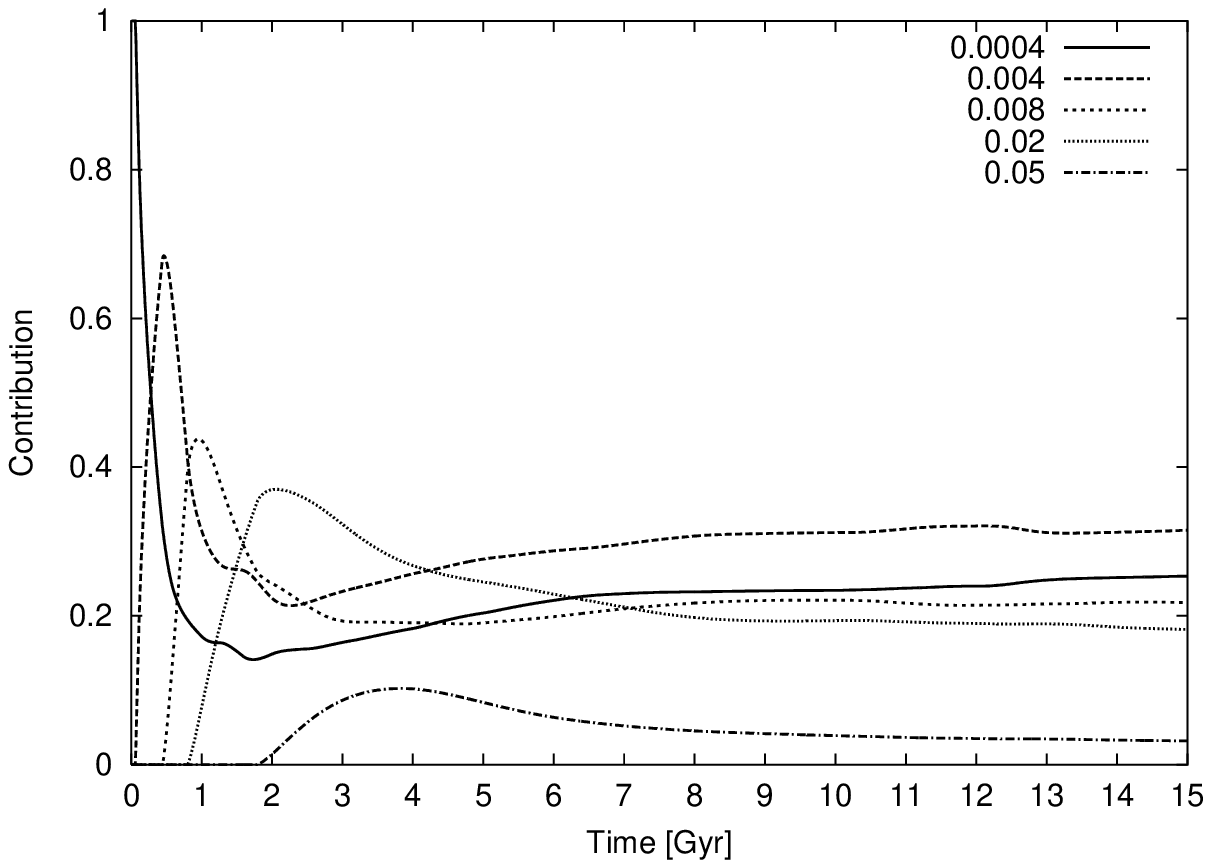}
\includegraphics[width=0.5\columnwidth]{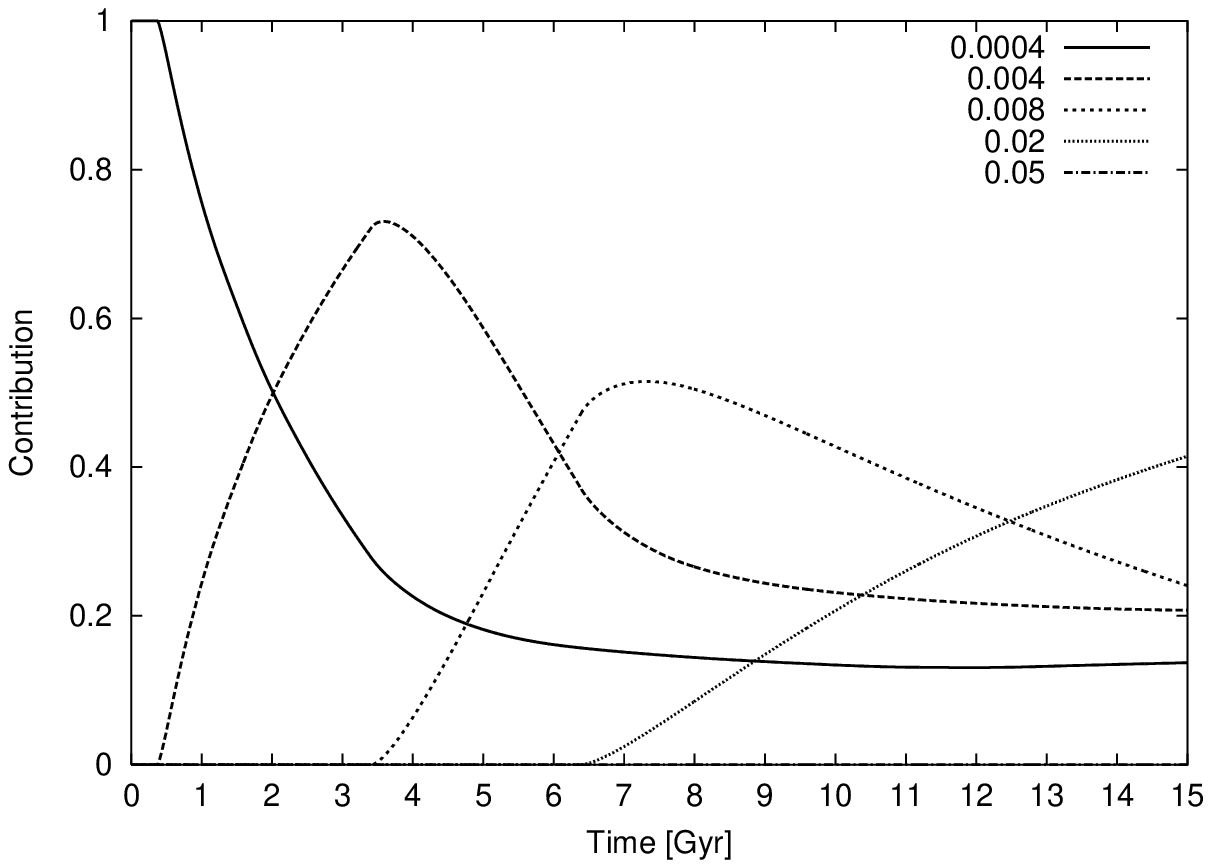}
\caption{Time evolution of the relative V-band luminosity contributions from stars of different metallicity subpopulations in E and Sb models.}
\end{figure}

\begin{figure}[ht]
\includegraphics[width=0.5\columnwidth]{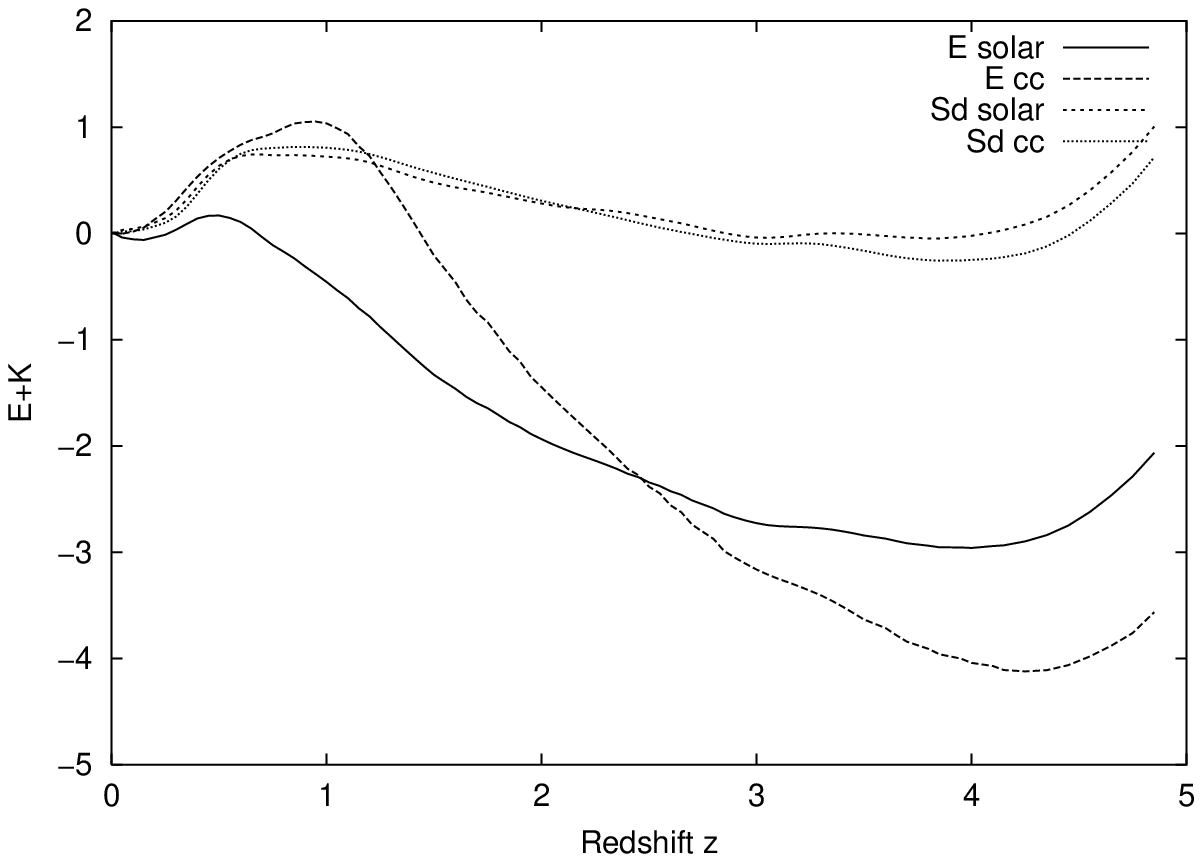}
\includegraphics[width=0.5\columnwidth]{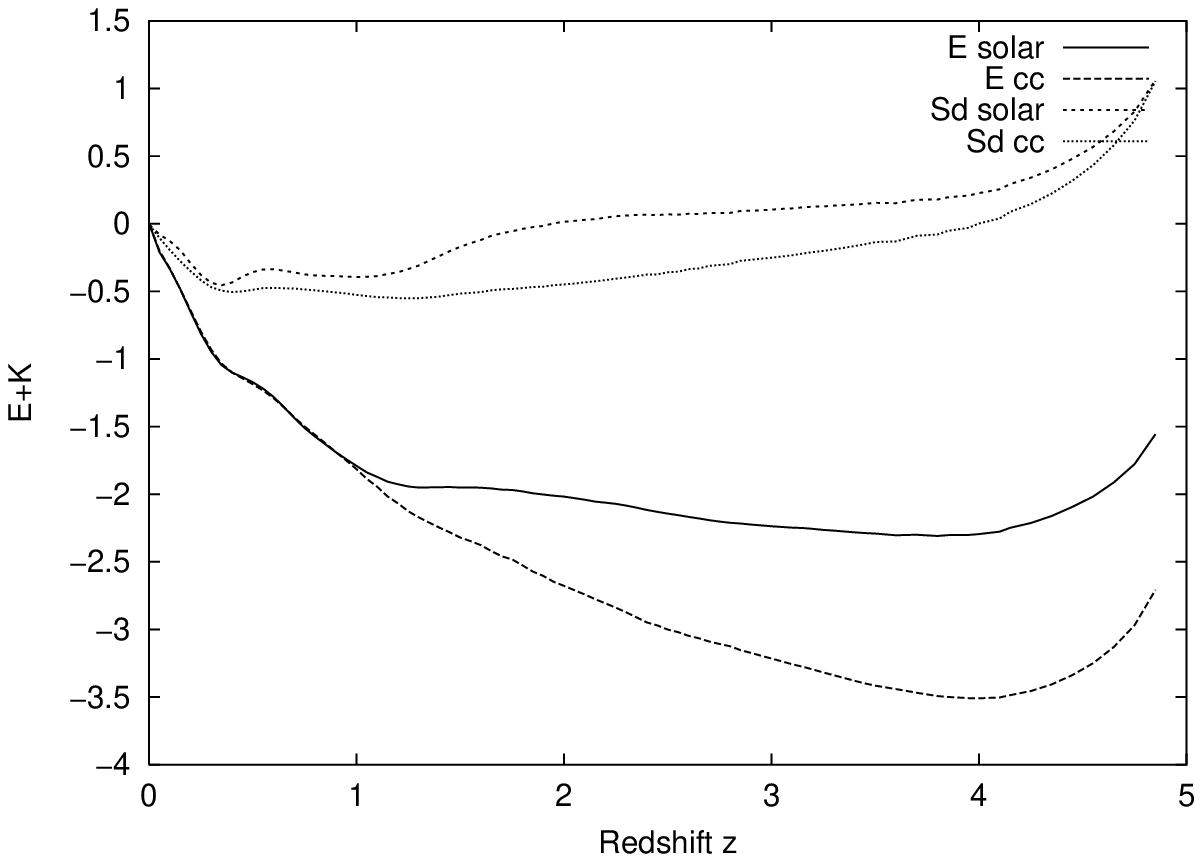}
\caption{Evolutionary and cosmological corrections in V (3a) and K (3b) for cc vs. solar metallicity E and Sb models.}
\end{figure}

Fig. 2 shows the time evolution of the relative luminosity contributions from stars of the 5 different metallicity subpopulations in the E and Sb models, respectively, to the light in V. Both models start with 100 \% of the V-band luminosity from the lowest metallicity stars. The broad stellar metallicity distribution of the E-model (Fig. 1a) is already established after $\sim 7$ Gyr and changes little thereafter. The Sb light is dominated by stars with subsolar metallicity during most of its evolution. Solar metallicity stars only start to contribute $\gta 20$ \% to the V-band luminosity after $\sim 10$ Gyr. As expected this affects the evolution of cosmological and evolutionary corrections as we show in Fig. 3 comparing the V- and K-band ${\rm (e+k)}$-corrections for cc E and Sd models with those of models using solar metallicity input physics only. These results point out the importance to include the ubiquitous contributions of low metallicity stars into spectro-cosmological galaxy evolution models. 

\vspace{-0.2cm}
\acknowledgements
P.W. and J.B. are partially supported by DFG grants 
Fr 916/6-2 and Fr 916/10-1 and gratefully acknowledge travel support from the organisers.

\end{article}

\vspace{-0.2cm}
\begin{thebibliography}{}
\bibitem[]{} Carollo, C. M., Danziger, I. J., 1994, MNRAS 270, 523 $+$ 743
\bibitem[]{} Ferguson, A. M. N., Gallagher, J. S., Wyse, R. F. G., 1998, AJ 116, 673
\bibitem[]{} Fritze -- v. Alvensleben, U., 1999, ASP Conf. Ser. 192, 273
\bibitem[]{} Grebel, E. K., 1997, Rev. Mod. Astron. 10, 29
\bibitem[]{} Kobayashi, C., Tsujimoto, T., \etal, 1998, ApJ 503, L155
\bibitem[]{} Lindner, U., Fritze - v. Alvensleben, U., Fricke, K. J., 1999, A\&A 341, 709
\bibitem[]{} Loewenstein, M., 1999, ASP Conf. Ser. 163, 153
\bibitem[]{} Lowenthal, J. D., Koo, D. C., Guzman, R., \etal, 1997, ApJ
481, 673
\bibitem[]{} Mehlert, D., Noll, S., Appenzeller \etal, 2002, A\&A 393, 809
\bibitem[]{} M\"oller, C. S., Fritze -- v. Alvensleben, U., Fricke, K. J., Calzetti, D., 2001, Ap\&SS 276, 799
\bibitem[]{} Pettini, M., Shapley, A. E., Steidel, C. C. \etal, 2001, ApJ 554, 981
\bibitem[]{} Pettini, M., Ellison, S. L., Bergeron, J., Petitjean, P., 2002, A\&A 391, 21
\bibitem[]{} Rocha - Pinto, H. J., Maciel, W. J., 1998, A\&A 339, 791
\bibitem[]{} Schulz, J., Fritze -- v. Alvensleben, U., M\"oller, C., Fricke, K. J., 2002, A\&A 392, 1
\bibitem[]{} Trager, S. C., Faber, S. M., Dressler, A., Oemler, A.,
1997, ApJ 485, 92
\bibitem[]{} van Zee, L., Salzer, J. J., Haynes, M. P., \etal, 1998, AJ 116, 2805
\end{thebibliography}
\end{document}